\documentstyle[12pt]{article}
\begin{document}

\begin{center}
{\large GAUGE INVARIANCE AND DUALITY IN THE NONCOMMUTATIVE PLANE }\\
\vskip 2cm
Subir Ghosh\\
\vskip 1cm
Physics and Applied Mathematics Unit,\\
Indian Statistical Institute,\\
203 B. T. Road, Calcutta 700108, \\
India.
\end{center}
\vskip 3cm
{\bf Abstract:}\\
We show that the duality between the self-dual and Maxwell-Chern-Simons theories in 2+1-dimensions survives when the space-time becomes noncommutative. Existence of the Seiberg-Witten map is crucial in the present analysis. It should be noted that the above models, being manifestly gauge variant and invariant respectively, transform differently under the Seiberg-Witten map. We also discuss this duality in the Stuckelberg formalism where the self-dual model is elevated to a gauge theory. The "`master"' lagrangian approach has been followed throughout.


\newpage
\section{Introduction}
\label{sec:Introduction} Lower dimensional gauge theories have
yielded unexpected subtleties in their structure, be it in the
form of massive propagating modes in the 1+1-dimensional Schwinger
model \cite {sj}, topologically massive gauge theories in
2+1-dimensions \cite{djt} or a "`hidden"' gauge invariance in
Self-Dual (SD) models in 2+1-dimensions \cite{djt,dj}. Once again,
as in the case of 1+1-dimensions, these bosonic models are
connected to fermionic models in the large fermion mass limit
\cite{fs}. Indeed, the above interrelations are interesting from
the point of view of providing unifications among diverse models
\cite{sg1} as well as exploiting the equivalences in explicit
computations.

Let us elaborate a little on the SD models in 2+1-dimensions as we will be concerned with them in the present work. Only abelian gauge theories will be considered here.
The local lagrangian formulation of them were first provided in \cite{tpn}. However, their equivalence with the well studied Maxwell-Chern-Simons (MCS) topological gauge theory \cite{djt} was revealed in \cite{dj}. Intuitively, the duality was expected since both SD and MCS models allow the propagation of a (parity violating) massive mode. But the above correspondence was clearly established by Deser and Jackiw \cite{dj} by the construction of a "`master"' lagrangian from which both SD and MCS models can be generated. The manifest gauge invariance of SD and the parent "`master"' model is {\it hidden} in the SD model, where the above symmetry is not manifest.

In recent years, Non-Commutative (NC) field theories \cite{rev}
and in particular NC Gauge Theories (NCGT) \cite{sw} have
generated a lot of interest due to their appearance in open string
and D-brane physics. This space-time noncommutativity modifies the
previous results in a non-trivial way and can lead to physically
interesting bounds on the NC parameter $\theta^{\rho\sigma}$. For
this reason a lot of effort is being given to re-derive existing
results in NC space-time.

In this paper, we are going to assess the effect of
noncommutativity in the above mentioned SD-MCS duality. This is
all the more pertinent since gauge invariance plays a major role
here and the noncommutativity affects gauge invariant and
non-invariant theories in strikingly different ways \cite{sw}. The
work of Seiberg and Witten \cite{sw} showed that there exists a
mapping, the Seiberg-Witten Map (SWM), that lifts a gauge theory
to its NC counterpart such that gauge orbits are mapped into NC
gauge orbits. The SWM is unique in the lowest nontrivial order of
$\theta^{\rho\sigma}$.  On the other hand, in a non-gauge theory,
noncommutativity affects only  products of fields (in the action),
without changing the individual field structures. This shows that
the SD and MCS theories will be generalized to NC field theories
in distinct ways due to their {\it manifest} gauge variance and
invariance respectively. We will explicitly demonstrate by way of
constructing the "`master"' lagrangian that in spite of the above
complications, {\it the SD-MCS duality is maintained in the NC
space-time}. Incidentally, the NCGT with abelian gauge group is
structurally quite akin to a non-abelian gauge theory in
conventional space-time. Hence our analysis might shed some light
on the SD-MCS duality in conventional space-time with a
non-abelian gauge group \cite{dj}. This constitutes the first part
of our work.

In the second part of our work, we show that there is another way of comparing noncommutative SD and MCS theories where the conventional SD theory is first embedded in a gauge invariant theory via the Stuckelberg formalism. Thus the starting SD and MCS theories, both being manifestly  gauge invariant, will be altered in similar ways under NC generalizations. We again establish the SD-MCS duality through the "`master"' lagrangian.

\section{Self-dual and Maxwell-Chern-Simons Duality in Noncommutative Space-time}
\label{sec:Self-dual and Maxwell-Chern-Simons Duality in NC Space-time}
The NC space-time is characterized by,
\begin{equation}
[x^{\rho},x^{\sigma}]_{*}=i\theta^{\rho\sigma}.
\label{nc}
\end{equation}
The $*$-product is given by the Moyal-Weyl formula,
\begin{equation}
p(x)*q(x)=pq+\frac{i}{2}\theta^{\rho\sigma}\partial_{\rho}p\partial_{\sigma}q+~O(\theta^{2}).
\label{mw}
\end{equation}
All our discussions will be valid up to the first non-trivial order in $\theta$.
The (inverse) SWM \cite{sw} to $O(\theta )$ is,
$$
A_{\mu}=\hat A_{\mu}-\theta^{\sigma\rho}\hat A_{\rho}(\partial_{\sigma}\hat A_{\mu}-\frac{1}{2}\partial_{\mu}\hat A_{\sigma})\equiv \hat A_{\mu}+\hat a_{\mu}(\hat A_{\nu},\theta )
$$
\begin{equation}
\lambda =\hat \lambda +\frac{1}{2}\theta^{\rho\sigma}\hat A_{\rho}\partial_{\sigma}\hat \lambda ,
\label{swm}
\end{equation}
where the "`hatted"' variables on the right live in NC space-time. The significance of the above map is that under an NC or $*$-gauge transformation of $\hat A_{\mu}$ by,
$$\hat\delta \hat A_{\mu}=\partial_{\mu}\hat \lambda +i[\hat\lambda ,\hat A_{\mu}]_{*},$$
$A_{\mu}$ will undergo the transformation $$\delta A_\mu =\partial_{\mu}\lambda $$ given in (\ref{tr}). Subsequently, under this mapping, a gauge invariant action in conventional space-time will be mapped to its NC counterpart, which will be $*$-gauge invariant.

Let us start by introducing the SD-MCS duality in conventional space-time.
For convenience we follow the notations  and metric $(g^{\mu\nu}=diag~(1,-1,-1))$ of \cite{dj}. The self-dual lagrangian, consisting of the ordinary and topological mass terms is \cite{tpn},
\begin{equation}
{\cal {L}_{SD}}=\frac{1}{2}f^\mu f_\mu-\frac{1}{2m}\epsilon ^{\alpha\beta\gamma}f_\alpha\partial _\beta f_\gamma .
\label{sd}
\end{equation}
On the other hand, the MCS model \cite{djt,dj} is described by,
\begin{equation}
{\cal {L}_{MCS}}=-\frac{1}{2}(\partial_{\alpha}A_{\beta}-\partial_{\beta}A_{\alpha})\partial^{\alpha}A^{\beta}+\frac{m}{2}\epsilon^{\alpha\beta\gamma}A_{\alpha}\partial_{\beta}A_{\gamma}.
\label{mcs}
\end{equation}
Note that total derivative terms in the lagrangian will be dropped throughout the present (classical) discussion. Clearly (\ref{mcs}) is invariant under the gauge transformation,
\begin{equation}
A_{\mu}\rightarrow A_{\mu}+\partial_{\mu}\lambda ,
\label{tr}
\end{equation}
whereas no such manifest symmetry exists for (\ref{sd}). Indeed, one can solve the equations of motion and constraints for both of the above models and show that there exists the identification \cite{dj},
$$f_{\mu}=\epsilon_{\mu\nu\tau}\partial^{\nu}A^{\tau}.$$
An alternative way is the construction of the following "`master"' lagrangian \cite{dj}
\begin{equation}
{\cal {L}_{DJ}}=\frac{1}{2}f^\mu f_\mu-\epsilon ^{\alpha\beta\gamma}f_\alpha\partial _\beta A_\gamma +\frac{m}{2}\epsilon^{\alpha\beta\gamma}A_{\alpha}\partial_{\beta}A_{\gamma},
\label{dj}
\end{equation}
which reduces to (\ref{sd}) or (\ref{mcs}) upon integrating out
$A_{\mu}$ or $f_{\mu}$ respectively, by exploiting the equations
of motion. Note that (\ref{dj}) is also gauge invariant.

The  mapping (\ref{swm}) together with $f_{\mu}=\hat f_{\mu}$ lifts the  "`master"' lagrangian in (\ref{dj}) to NC "`master"' lagrangian,
$$
\hat{\cal {L}}_{DJ}=\frac{1}{2}\hat f^\mu \hat f_\mu-\epsilon ^{\alpha\beta\gamma}\hat f_\alpha\partial _\beta (\hat A_\gamma +\hat a_{\gamma}) +\frac{m}{2}\epsilon^{\alpha\beta\gamma}(\hat A_{\alpha}+\hat a_{\alpha})\partial_{\beta}(\hat A_{\gamma}+\hat a_{\gamma})$$
\begin{equation}
=\frac{1}{2}\hat f^\mu \hat f_\mu -\epsilon ^{\alpha\beta\gamma}\hat f_\alpha\partial _\beta (\hat A_\gamma +\hat a_{\gamma})+\frac{m}{2}\epsilon^{\alpha\beta\gamma}(\hat A_{\alpha}\partial_{\beta}\hat A_{\gamma}+2\hat a_{\alpha}\partial_{\beta}\hat A_{\gamma})+~O(\theta^{2}).
\label{ncdj}
\end{equation}
Let us now follow the computational scheme, (of selective integration of fields), used in the conventional space-time. Variation of $\hat f_{\mu}$ yields the equation,
\begin{equation}
\hat f_{\mu}=\epsilon_{\mu\nu\tau}\partial^{\nu}(\hat A^{\tau}+\hat a^{\tau}).
\label{f}
\end{equation}
Substituting $\hat f_{\mu}$ from (\ref{f}) in (\ref{ncdj}), we obtain,
\begin{equation}
\hat{\cal {L}}_{DJ}=-\frac{1}{2}(\partial_{\mu}\hat A_{\nu}-\partial_{\nu}\hat A_{\mu})\partial^{\mu}\hat A^{\nu}-(\partial_{\mu}\hat A_{\nu}-\partial_{\nu}\hat A_{\mu})\partial^{\mu}\hat a^{\nu}+\frac{m}{2}\epsilon_{\mu\alpha\beta}(\hat A^{\mu}\partial^{\alpha}\hat A^{\beta}+2\hat A^{\mu}\partial^{\alpha}\hat a^{\beta}).
\label{ncmcs}
\end{equation}
To $O(\theta )$, this is nothing but $\hat{\cal {L}}_{MCS}$ where
\begin{equation}
\hat{\cal {L}}_{MCS}=-\frac{1}{2}[\partial_{\alpha}(\hat A_{\beta}+\hat a_\beta ) -\partial_{\beta}(\hat A_{\alpha}+\hat a_\alpha )][\partial^{\alpha}(\hat A^{\beta}+\hat a_\beta )]+\frac{m}{2}\epsilon^{\alpha\beta\gamma}(\hat A_{\alpha}+\hat a^\alpha )\partial_{\beta}(\hat A_{\gamma}+\hat a_\gamma ).
\label{nmcs}
\end{equation}
In a similar manner, variation of $\hat A_{\mu}$ reproduces the equation of motion,
\begin{equation}
m\epsilon_{\mu\alpha\beta}\partial^{\alpha}\hat A^{\beta}\approx \epsilon_{\mu\alpha\beta}\partial^{\alpha}(\hat f^{\beta}-m\hat a^{\beta})-\epsilon_{\nu\alpha\beta}(m\partial^{\alpha}\hat A^{\beta}-\partial^{\alpha}\hat f^{\beta})\frac{\delta \hat a^{\nu}}{\delta \hat A^{\mu}} .
\label{a}
\end{equation}
Notice that the last term in the right hand side of (\ref{a}) is actually of $O(\theta^{2})$ and hence can be omitted. Once again putting back the above relation (\ref{a}) in (\ref{ncdj}) we recover $\hat{\cal {L}}_{SD}$,
\begin{equation}
\hat{\cal {L}}_{SD}=\frac{1}{2}\hat f^\mu \hat f_\mu-\frac{1}{2m}\epsilon ^{\alpha\beta\gamma}\hat f_\alpha\partial _\beta \hat f_\gamma .
\label{nsd}
\end{equation}
This constitutes the demonstration that (to $O(\theta )$) the SD-MCS duality remains intact in noncommutative space-time even though the mappings of the SD and MCS models to their respective noncommutative counterparts are entirely different. It appears that, to the first non-trivial order in $\theta $ (that we are presently interested in), existence of the perturbative form of the SWM \cite{sw} is sufficient to guarantee the SD-MCS duality in noncommutative space-time, irrespective of the explicit structure of the map in (\ref{swm}). We will return to this point briefly at the end.
\section{Duality in the Stuckelberg Formalism}
\label{sec:Duality in the Stuckelberg Formalism}

For various reasons it sometimes turns out to be convenient to embed the particular model under study in a larger phase space in which the extended model has a gauge invariance. The original model and its gauge invariant extension are obviously gauge equivalent so that the physical ({\it i.e.} gauge invariant) sector of the original model remains intact. Stuckelberg formalism is one of the simplest such prescription where the gauge variant SD model ${\cal {L}_{SD}}$ in (\ref{sd}) can be extended to the manifestly gauge invariant ${\cal {L}_{SD}^S}$,
\begin{equation}
{\cal {L}_{SD}^S}=\frac{1}{2}(f^\mu -\partial^{\mu}\phi )( f_\mu-\partial_{\mu}\phi)-\frac{1}{2m}\epsilon ^{\alpha\beta\gamma}f_\alpha\partial _\beta f_\gamma .
\label{sds}
\end{equation}
Here $\phi$ is an auxiliary degree of freedom, which is introduced to induce the invariance of ${\cal {L}_{SD}^S}$ under the gauge transformation,
\begin{equation}
f_{\mu}\rightarrow f_{\mu}+\partial_{\mu}\sigma ~,~\phi \rightarrow \phi + \sigma .
\label{gi}
\end{equation}
Obviously $\phi$ is a gauge degree of freedom and the gauge choice $\phi =0$ reproduces ${\cal {L}_{SD}}$.

The "`master"' lagrangian ${\cal {L}_{DJ}^S}$ corresponding to the pair, ${\cal {L}_{SD}^S}$ of (\ref{sds}) and the earlier MCS model (\ref{mcs}), (which remains unchanged), is given by the following extended version of ${\cal {L}_{DJ}}$,
\begin{equation}
{\cal {L}}_{DJ}^S=\frac{1}{2}(f^\mu -\partial^{\mu}\phi ) (f_\mu-\partial_{\mu }\phi )-\epsilon ^{\alpha\beta\gamma}f_\alpha\partial _\beta A_\gamma +\frac{m}{2}\epsilon^{\alpha\beta\gamma}A_{\alpha}\partial_{\beta}A_{\gamma}.
\label{st}
\end{equation}
The above model (\ref{st}) actually has two independent gauge invariances,
\begin{equation}
A_{\mu}\rightarrow A_{\mu}+\partial_{\mu}\lambda ~;~f_{\mu}\rightarrow f_{\mu}+\partial_{\mu}\sigma ~,~\phi \rightarrow \phi + \sigma .
\label{ggi}
\end{equation}
Now we consider the SWM appropriate for the present gauge theory (\ref{st}) having two gauge invariances. Thus, the SWM of (\ref{swm}) has to be augmented by the following map,
$$
f_{\mu}=\hat f_{\mu}-\theta^{\alpha\rho}\hat f_{\rho}(\partial_{\alpha}\hat f_{\mu}-\frac{1}{2}\partial_{\mu}\hat f_{\alpha})\equiv \hat f_{\mu}+\hat h_{\mu}(\hat f_{\nu},\theta )
$$
\begin{equation}
\sigma =\hat \sigma +\frac{1}{2}\theta^{\rho\alpha}\hat f_{\rho}\partial_{\alpha}\hat \sigma \equiv \hat \sigma +\hat \sigma ^1,
\label{swm1}
\end{equation}
The full SWM leads to the NC counterpart of ${\cal {L}}_{DJ}^S $,
$$
\hat
{\cal {L}}_{DJ}^S=\frac{1}{2}(\hat f^\mu +\hat h^\mu -\partial^{\mu}\hat\phi ) (\hat f_\mu +\hat h_\mu-\partial_{\mu }\hat\phi )-\epsilon ^{\alpha\beta\gamma}(\hat f_\alpha +\hat h_{\alpha })\partial _\beta (\hat A_\gamma +\hat a_\gamma )$$
$$+\frac{m}{2}\epsilon^{\alpha\beta\gamma}(\hat A_{\alpha}+\hat a_\alpha )\partial_{\beta}(\hat A_{\gamma}+\hat a_\gamma )$$
$$=\frac{1}{2}(\hat f^\mu -\partial^{\mu} \hat \phi )( \hat f_\mu -\partial_{\mu}\hat\phi )
+(\hat f^\mu -\partial^{\mu} \hat \phi )\hat h_\mu
-\epsilon ^{\alpha\beta\gamma}(\hat f_\alpha\partial _\beta \hat A_\gamma +
\hat f_\alpha\partial _\beta \hat a_{\gamma}+\hat h_\alpha\partial _\beta \hat A_{\gamma})$$
\begin{equation}
+\frac{m}{2}\epsilon^{\alpha\beta\gamma}(\hat A_{\alpha}\partial_{\beta}\hat A_{\gamma}+2\hat a_{\alpha}\partial_{\beta}\hat A_{\gamma})+~O(\theta^{2}).
\label{st1}
\end{equation}
Two points regarding this model are to be noted. Firstly, although there are two independent
gauge invariances with distinct gauge parameters $\lambda$ and $\sigma$, this feature does not
show up in $\hat{\cal {L}}_{DJ}^S$, since the SWM for the gauge field is independent of the
gauge transformation function.

Secondly, from the basic tenet of the SWM, that for abelian gauge group, gauge invariant
sectors in ordinary spacetime under SWM will be translated to *-gauge invariant sectors in
NC spacetime, we conclude that the combination $\hat f^\mu +\hat h^\mu -\partial ^\mu \hat \phi $ has to be *-gauge invariant, since $f^\mu -\partial ^\mu\phi $ is gauge invariant. This shows that under *-gauge transformation,
\begin{equation}
\hat\delta\hat\phi =\hat\sigma -\frac{1}{2}\theta^{\alpha\beta}\hat f_{\beta}\partial_{\alpha}\hat\sigma .
\label{**}
\end{equation}
Now, we follow the formal way \cite{suo} of extracting the SWM for $\phi$ and use the
following expansions,
\begin{equation}
\hat\phi =\phi +\varphi (\theta ) +O(\theta^{2})~~;~\hat f^{\mu}=f^{\mu}+h^{\mu}(\theta )
+O(\theta^{2})~~;~\hat\sigma =\sigma +\sigma_{1}(\theta )+O(\theta^{2}),
\label{*}
\end{equation}
in (\ref{**}). $\hat h^\mu $ and $\sigma _1$ are defined in
(\ref{swm1}). Comparing powers of $\theta$, we get
\begin{equation}
\delta\phi =\sigma ~~;~~\delta\varphi =0.
\label{***}
\end{equation}
But such a gauge invariant structure for $\varphi$ is impossible to construct indicating that $\varphi =0$ and so $\hat \phi =\phi +O(\theta^{2})$.

Proceeding in exactly the same way as before, the variational equations of motion are obtained as,
\begin{equation}
 \hat f_{\mu}-\partial_{\mu}\hat\phi =\epsilon_{\mu\alpha\beta}\partial^{\alpha}\hat A^{\beta}+\epsilon_{\mu\alpha\beta}\partial^{\alpha}\hat a^{\beta}-\hat h_{\mu}
+(\epsilon_{\nu\alpha\beta}\partial^{\alpha}\hat A^{\beta}-\hat f^{\nu}+\partial^{\nu}\hat\phi )\frac{\delta \hat h_{\nu}}{\delta\hat f^{\mu}} +O(\theta^{2}),
\label{f1}
\end{equation}
\begin{equation}
m\epsilon_{\mu\alpha\beta}\partial^{\alpha}\hat A^{\beta}=\epsilon_{\mu\alpha\beta}\partial^{\alpha}(\hat f^{\beta}-m\hat a^{\beta}+\hat h^{\beta})-\epsilon_{\nu\alpha\beta}(m\partial^{\alpha}\hat A^{\beta}-\partial^{\alpha}\hat f^{\beta})\frac{\delta \hat a^{\nu}}{\delta \hat A^{\mu}}+O(\theta^{2}) .
\label{a1}
\end{equation}
Note that in both the above equations (\ref{f1}) and (\ref{a1}), the last terms in the right hand side containing variational derivatives are actually of $O(\theta )^2$ and hence can be left out in the present analysis.
Substitution of (\ref{f1}) in (\ref{st1}) reproduces (\ref{nmcs}), the noncommutative version of the MCS model. On the other hand, using (\ref{a1}) in (\ref{st1}) we get, $\hat{\cal {L}}_{SD}^S$, the noncommutative analogue of ${\cal {L}}_{SD}^S$ in (\ref{sds}),
\begin{equation}
\hat{\cal {L}}_{SD}^S=\frac{1}{2}(\hat f^\mu +\hat h^\mu -\partial^{\mu} \hat \phi )( \hat f_\mu +\hat h_\mu -\partial_{\mu}\hat\phi )-\frac{1}{2m}\epsilon^{\alpha\beta\gamma}(\hat f_{\alpha}\partial_{\beta}\hat f_{\gamma}+2\hat h_{\alpha}\partial_{\beta}\hat f_{\gamma}).
\label{sd1}
\end{equation}
This exercise shows that the DS-MCS duality remains intact in NC
space-time, when the SD model is also treated as a gauge theory in
the Stuckelberg formalism.

It should be mentioned that from the variational nature of the problem, it might seem that
under any change of the field variable to the lowest non-trivial order, the equivalence or duality relations of this sort will remain intact. However, in general this may not be true. The explicit nature of the "`master"' lagrangian as well as the induced lagrangians, and the fact that the Seiberg-Witten map does not distinguish between different gauge invariances, (as far as the field variables are concerned), all contribute to the above derivations.

It would be very interesting to establish the equivalence between self dual and Maxwell-Chern-Simons theories for arbitrary orders in $\theta $, since that will bring into play the explicit forms of the SWM. However, that has not been included in the present work mainly because (I): $O(\theta)$ computations and results have a very important role in NC theories, and (II): there are ambiguities \cite{theta} in the SWM for higher orders in $\theta$. On the other hand, the above duality may be exploited to favour one expression of the SWM over the other alternatives for say $O(\theta^{2})$. These issues will be addressed separately.
\vskip .2cm
{\bf Note addeded:}\\
It is straightforward to show that the duality between Maxwell-Chern-Simons and self dual theories, in the noncommutative plane, holds to {\it all} orders in $\theta $. We start with the full equation corresponding to (\ref{a}),
\begin{equation}
\epsilon^{\sigma\beta\gamma}\partial_{\beta}(\hat f-m\hat A)_{\gamma}(g^{\alpha}_{\sigma}+
\frac{\delta \hat c_{\sigma}}{\delta \hat A_{\alpha}})=m
\epsilon^{\sigma\beta\gamma}\partial_{\beta}\hat c_{\gamma}(g^{\alpha}_{\sigma}+
\frac{\delta \hat c_{\sigma}}{\delta \hat A_{\alpha}}),
\label{aa}
\end{equation}
where $\hat c^{\mu}$ denotes the full $\theta $ correction to $A^{\mu}$. Since the operator $(g^{\alpha}_{\sigma}+
\frac{\delta \hat c_{\sigma}}{\delta \hat A_{\alpha}})$ is invertible, one gets the exact relation
\begin{equation}
\epsilon^{\sigma\beta\gamma}\partial_{\beta}(\hat f-m\hat A)_{\gamma}=m
\epsilon^{\sigma\beta\gamma}\partial_{\beta}\hat c_{\gamma}.
\label{aaa}
\end{equation}
Substituting this in the NC master Lagrangian (\ref{ncdj}) (with $\hat a^{\mu}$ replaced by $\hat c^{\mu}$), reproduces the self dual model.

\vskip .5cm
{\bf Acknowledgement:} It is a pleasure to thank the referee for the constructive comments.

\end{document}